# Design of Fast Response Smart Electric Vehicle Charging Infrastructure

Ching-Yen Chung, Joshua Chynoweth, Charlie Qiu, Chi-Cheng Chu, Rajit Gadh
Department of Mechanical and Aerospace Engineering
University of California, Los Angeles
Los Angeles, USA
chingyenchung@ucla.edu, joshuachynoweth@gmail.com,
qiucharlie@gmail.com, peterchu@ucla.edu, gadh@ucla.edu

*Abstract*—The response time of the smart electrical vehicle (EV) charging infrastructure is the key index of the system performance. The traffic between the smart EV charging station and the control center dominates the response time of the smart charging stations. To accelerate the response of the smart EV charging station, there is a need for a technology that collects the information locally and relays it to the control center periodically. To reduce the traffic between the smart EV charger and the control center, a Power Information Collector (PIC), capable of collecting all the meters' power information in the charging station, is proposed and implemented in this paper. The response time is further reduced by pushing the power information to the control center. Thus, a fast response smart EV charging infrastructure is achieved to handle the shortage of energy in the local grid.

## I. INTRODUCTION

Electrical Vehicles (EV) are becoming a viable choice for many people as the world is beginning to lower its dependence on fossil fuels. Charging stations in parking structures and garages will become important for longer distance commuters as more EVs are being purchased and used. They will now have to charge during on peak hours which will stress the grid. In order to satisfy the demand of EV charging requests, an EV charging management system needs to be implemented to handle the peak demand of EV charging, regardless whether charging takes place in a parking garage or at home. The need for a reliable infrastructure for monitoring and controlling vehicle charging is a top priority. When there is a power shortage, charging stations may choose to turn off the power or reduce the current to the EVs in order to lower the impact on the power grid. As long as the total power consumed is within the safety limit of a given circuit, smart EV charging stations should be capable of charging several EVs from that single circuit. This charging should be done with different duty cycles and power levels. UCLA Smart-Grid Energy Research Center (SMERC) has developed WINSmartEV[TM][1][2][3] a software based EV charging system that includes monitoring, control, and management for EV chargers. This system also includes multiplexed charging stations capable of providing varying power to several EVs from one circuit. With this smart charging infrastructure, shortages of energy in local grids can be prevented by managing EV charging currents and sessions.

The existing WINSmartEV[TM] collects the power information from the meters by using a data pull method. Because wireless infrastructure is available and ubiquitous, the chargers are generally connected to the internet through 3G cellular networks. Local communication between and within chargers is through ZigBee mesh network. Therefore, when the data pull method sends a power information request command from the server to a charger, the signal must pass through the internet and through a 3G network before it reaches the gateway of the charging station. Then the gateway relays the power information request command to the specific meter it is meant for. When the gateway receives a reply from the meter, it relays the response back to the server where the information travels back in reverse order. With multiple meters requiring multiple requests each, the aggregated round trip times cause slow performance. In order to enhance the system's performance and shorten the response time of the system, a device is proposed that will collect the power information locally in order to send it in to the server together as one packet. By decreasing the number of communications required for status reports and control operations, this Power Information Collector (PIC) will significantly decrease the delay time for switching EV charging sessions or changing current to the EVs. By reducing the traffic between the server and the charging stations, the proposed improvements allow the control center to serve a larger system, which enhances the capability of the existing WINSmartEV[TM] framework. Moreover, because the charging station's local control unit has access to the power information of the meters, it can employ local charging algorithms to control charging. This will leave the control server with more computation power to handle a larger system.

Current commercial EV charging stations, Leviton[4] and Clippercreek[5], simply provide basic charging stations without power information monitoring and network features. Neither remote nor local charging algorithms are implemented. Since there is no smart charging algorithm running on their charging stations, there is no need for fast response devices. Other commercial charging stations like Coulomb[6] and Blink[7] have their own proprietary network and are likely to have smart charging algorithms on their charging stations. Because these charging stations only have one or two outlets, which are not for current sharing purpose, the power information collections would have less effect on these stations. In [8][9][10][11], although several charging algorithms and simulations are presented, none of them discuss how to improve the performance of the smart charging system.

The paper is structured in the following way: first, the existing WINSmartEV[TM] architecture is introduced. Second, the architecture of the proposed system is presented and details of the Power Information Collector (PIC) are described. Last, the results of the experiments are presented and discussed.

## II. EXISTING WINSMARTEV[TM] SYSTEM

The existing WINSmartEV[TM] smart charging infrastructure including the network architecture and the design of the smart charging station is introduced in this section. A four-channel WINSmartEV[TM] smart charging station in a UCLA parking structure is shown in Fig. 1.

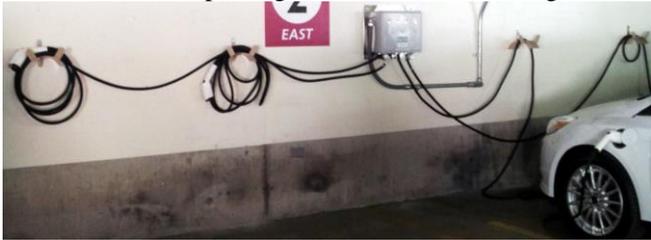

Figure 1. WINSmartEV[TM] 4-channel smart charging station

Multiple EVs can be charged with variable-current at one time. The authorized users can check the availability of charging stations, start or stop EV charging, check the charging status, view monthly charging status, and manage user account with user's mobile device. Fig. 2 shows the screen shot of the mobile app of User Control Center on user's mobile device.

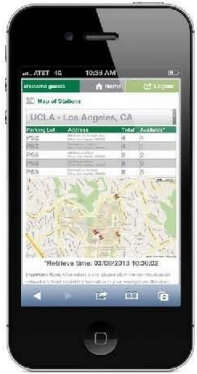

Figure 2. Screen shots of User Control Center

The network architecture of the WINSmartEV[TM] smart charging infrastructure is shown in Fig. 3.

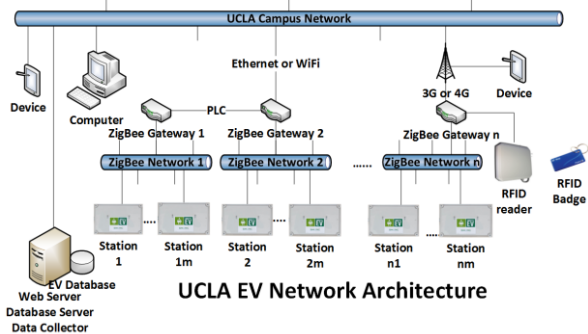

Figure 3. WINSmartEV[TM] Infrastructure

All charging stations are controlled by a server-based aggregate charging system through multiple protocol gateways with 3G connection. The system is applicable wherever a cellular signal exists, especially where wired or WiFi communication is unavailable. Fig 4 details the communication method within a charging station.

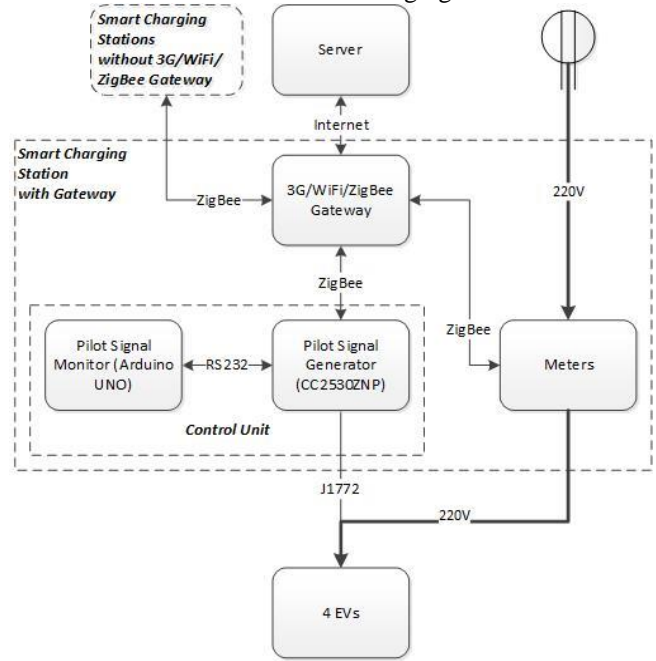

Figure 4. Details of smart charging station

The communication between the gateway, meters, and the control unit is established by using ZigBee mesh network. In a given geographic location, a gateway-equipped master station and all other non-gateway–equipped stations also communicate each other by utilizing ZigBee mesh network.

To monitor and control the charging activities, the server-based aggregate charging control system is employed. Four major software components serve the system including: Database, Station Controller and Data Collector, System Monitor and Control Center, and User Control Center. The Station Controller and Data Collector send commands to the charging station to both control the charging and collect the meters' power information. The charging stations can be manually controlled by an administrator through the Monitor and Control Center shown in Fig. 5.

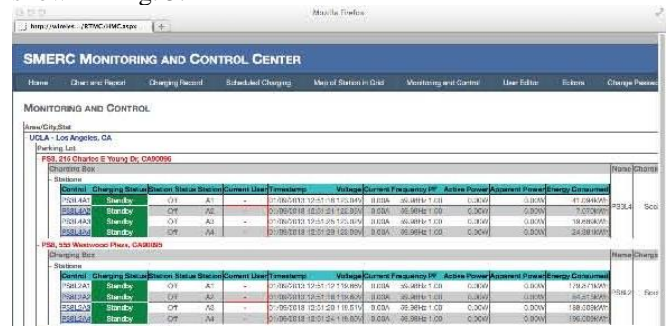

Figure 5. Screenshot of Monitoring and Control Center

## III. PROPOSED POWER INFORMATION COLLECTOR (PIC)

For the server to obtain data from the meters in the existing system, each request must pass through multiple stages before the information is sent back. The requests begin as HTTP commands sent out by the server over the internet and through the 3G network before reaching the gateway. The gateway then relays the commands through the Zigbee mesh network to the meters. Each response travels the same path in reverse direction. Because there are 2 sets of data, power information and ON/OFF status, required in each meter update, two commands are required to update each meter. Because there are 4 meters in each charging station, 8 of these HTTP commands are required to update the data for one charger.

In order to change 8 requests to 1 request for each status update, and thereby cutting the time by a factor of 8, a local data collector, Power Information Collector (PIC), is proposed in this section. This PIC will collect and compile the entire list of information on a 4-channel charging station that the server requires and send it all upon one request. This will reduce 3G traffic by cutting the request from required for a status update from 8 to 1. Because the collector will connect directly to the gateway, the HTTP request never has to pass through the ZigBee mesh network. Therefore, the response time for the one request is shorter than the response time for each of the 8 requests required to update the data on the current system. Because the data collector can cycle its updates quickly, the information that the control server receives will always be up to date. The smart charging station with PIC is shown in Fig. 6.

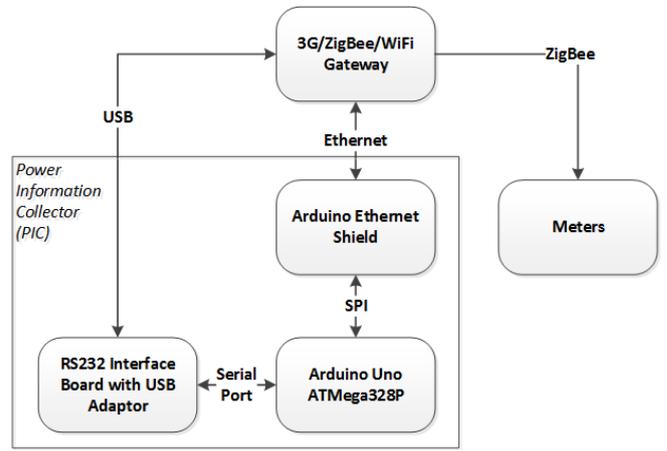

Figure 7. Schematic of Power Information Collector (PIC)

The PIC consists of an Arduino UNO board with ATMega328P microprocessor, an Arduino Ethernet Shield, and a RS232 serial port interface board. RS232 serial port interface board and the Arduino Ethernet Shield are wired to the gateway for communication. The RS232 serial port communicates between the server and the PIC. The Ethernet port retrieves the power information and ON/OFF status of the meters. Fig. 8 shows the firmware flow on the ATMega328P microprocessor of the PIC.

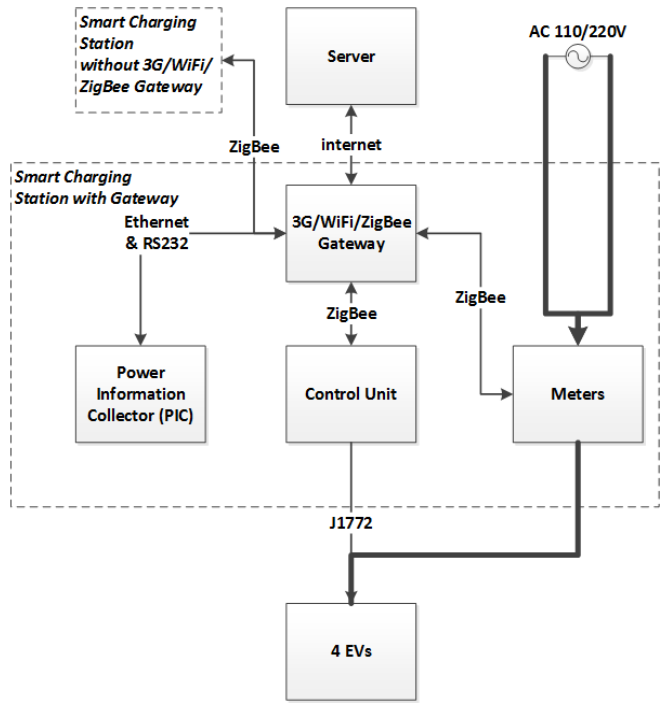

Figure 6. Smart Charging station with Power Information Collector (PIC)

Fig. 7 shows the schematic of the PIC and the connections between the multiple protocol gateway.

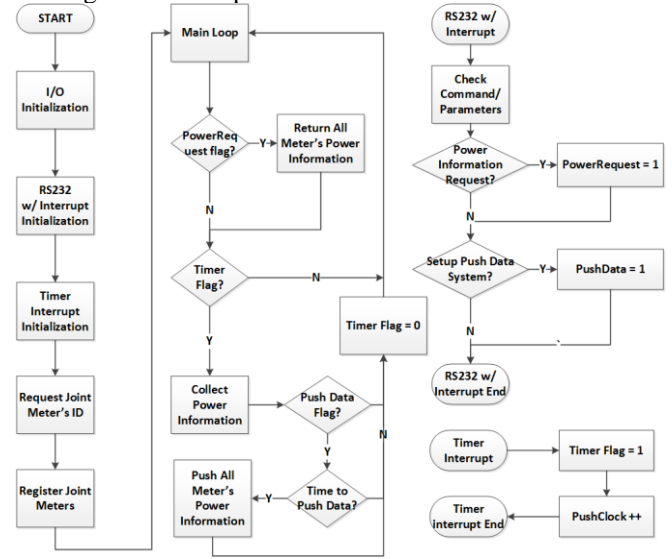

Figure 8. Firmware Flow in the PIC

There are three loops in the firmware flow including the main loop, RS232 interrupt loop, and timer interrupt loop. Before the main loop is the startup initialization. The RS232 with internal interrupt and timer interrupt are initialized at the startup process. After that, the meters' ID are requested and registered for the power information and relay status retrievals later. The commands from the gateway include the power information request and the pushing data system setup. These commands are handled by the RS232 within the interrupt loop; thus, commands will not be missed while executing other processes in the main loop. In the RS232 interrupt loop, only command checks and flag sets are handled; most of the actions are handled in the main loop according to these flags. With the timer interrupt, the power information and the status of the relays can be pushed to the control center periodically by setting up the PushData flag. The period of data-pushing can be changed remotely, which makes the service more flexible. To avoid unexpected actions, the time of the timer interrupt routine should be as small as possible. Therefore, in the timer interrupt routine, only the timer flag is handled to avoid missing any RS232 commands or breaking a timer interrupt loop. Note that the time required for the power information collection process and the data pushing process should be less than the interval of the timer interrupts so that these processes can be handled correctly. The hardware implementation of the PIC is shown in Fig. 9

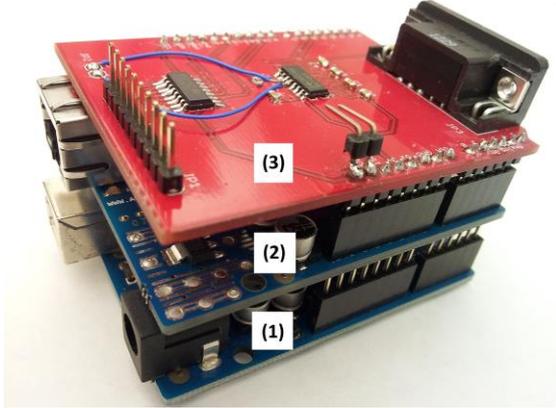

Figure 9. Implementation of Power Information Collector: (1) Microprocessor ATMega328P (2) Ethernet Shield (3) RS232 Interface with Relay Driver

## IV. EXPERIMENTS AND RESULTS

In this section, the distributions of round trip time of power information retrieval are measured and presented. The round trip time of power information retrieval corresponds to the waiting time of the server for the response from the charging station after sending out the power information retrieval command. The round trip time, $T_{RoundTrip}$, is expressed in (1):

$$T_{RoundTrip} = T_{Network} + T_{Metering}$$
$$= (T_{Server\_Cloud} + T_{Cloud} + T_{Cloud\_Station}) + T_{Metering} \quad (1)$$

$T_{Metering}$ stands for the time required by the meter in retrieving the power information and it is roughly estimated in the experiments. $T_{Network}$ includes the time required between the server and the charging station. The charging station can connect to the internet through different networks including Ethernet, WiFi, and 3G, which makes $T_{Cloud\_Station}$ the dominant factor in the round trip time of power information between the server and the charging station, $T_{RoundTrip}$. Fig. 10 shows the distribution of the round trip times. The round trip times were recorded every 5 minutes in one week under different network configurations. The upper chart is with Ethernet while the lower chart is with WiFi. Here we use $T_{Ethernet}$ and $T_{WiFi}$ for $T_{Cloud\_Station}$ in these two cases. Note that the server and the charging station are under the same default gateway in the experiment.

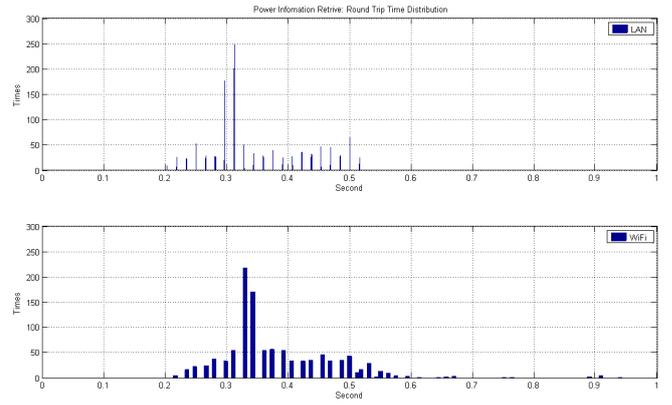

Figure 10. $T_{RoundTrip}$ with Ethernet(Upper) and WiFi(Lower)

The experiment shows that the shortest $T_{RoundTrip}$ is around 0.2 seconds. Normally, when the server and the charging station are under the same default gateway with Ethernet, the $T_{Ethernet}$ is within the microsecond degree, which could be neglected in $T_{RoundTrip}$. $T_{Cloud}$ is also approximate to 0 in this case; therefore, $T_{RoundTrip}$ will be approximate to $T_{Meterging}$, which is expressed in (2):

$$T_{RoundTrip\_Ethernet} = (T_{Ethernet} + T_{Cloud}) + T_{Metering} \approx T_{Metering} \quad (2)$$

Here, $T_{Metering}$ is the dominant factor in Ethernet case; therefore, the distribution of $T_{Metering}$ is approximate to $T_{RoundTrip}$ in Ethernet case. Comparing with Ethernet case, the lower chart of Fig.10 shows that $T_{WiFi}$ is slightly slower than $T_{Ethernet}$. Note that $T_{Metering}$ is still the dominant factor in WiFi case.

As for the 3G case, $T_{RoundTrip}$ distributions of 3 charging stations in different locations: UCLA, LA downtown, and Santa Monica are measured in second, third, fourth charts in Fig. 11. The first chart of Fig. 11 is the Ethernet case for comparison.

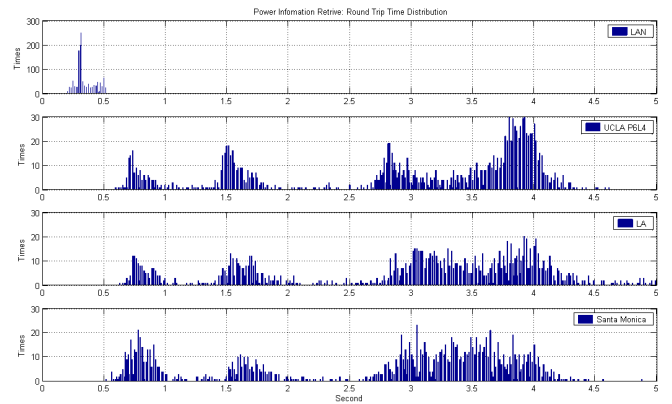

Figure 11. $T_{RoundTrip}$ with 3G in 3 different locations

The round trip time $T_{RoundTrip}$ of 3G cases can be expressed in (3):

$$T_{RoundTrip\_3G} = (T_{Server\_Cloud} + T_{Cloud} + T_{3G}) + T_{Metering} \approx T_{3G} + T_{Metering} \quad (3)$$

The experiments shows that $T_{3G}$ is not a constant offset from Ethernet case but a distribution with probability with four peaks measured in its distribution. The distribution of round trip time in 3G cases, $T_{RoundTrip\_3G}$, shows no apparent difference in different locations; therefore, the round trip time model could be used in other places as long as they have the same configuration.

Considering that $T_{3G}$ could be time variant during the day and the week, Fig. 12 shows $T_{RoundTrip\_3G}$ during the week and the days of the charging station at UCLA. The first chart of Fig. 12 is $T_{RoundTrip\_3G}$ during a week and the x-axis of first chart stands for seven days. (0 to 1 is Sunday, 1 to 2 is Monday, etc.) The second and the third charts are the $T_{RoundTrip\_3G}$ on Sunday and Monday.

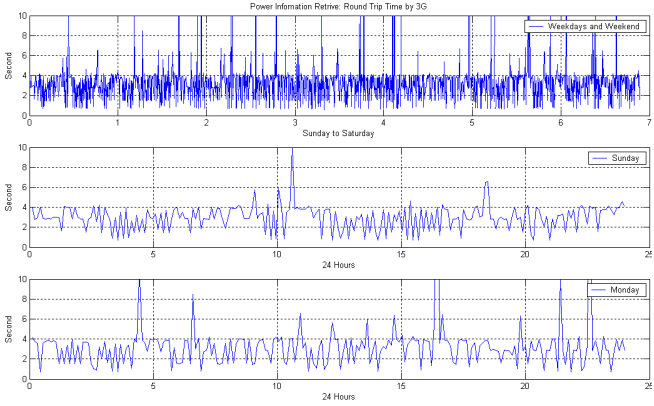

Figure 12. $T_{RoundTrip\_3G}$ during the week and the day.

The results show that $T_{RoundTrip\_3G}$ are faster during certain hours of Sunday and Monday, but the range of the distribution is no different in these two cases.

## V. DISCUSSION

### A. Improvement on System's Response Time

In order to retrieve data from a charging station's four meters in the 3G case of the data pull system, from (3), , the control center needs to wait:

$$T_{Retrieval} = 4T_{RoundTrip\_3G} \approx 4(T_{3G} + T_{Metering}) \quad (4)$$

From the experiment in section III, when we take the worst cases into account, the control center needs to wait 20 seconds to finish power information retrievals. With PIC, only one time request is require and $T_{Metering}$ is eliminated by its periodical retrievals. This leaves only time $T_{3G}$ for the server to retrieve the power information. Because $T_{3G}$ is 4.5 seconds in the worst case, the new PIC system is 4.4 times faster than the original system. The new PIC system is 8.4 times faster when taking the ON/OFF status retrieval into account.

The system performance can be further improved by setting the PIC to be a data pushing device, which means PIC periodically sends the power information from the meters to the control center. The control center can have the meters' power information without waiting, which greatly improve the performance of the system. One concern here is that the power information is not real time data. From the experiments in section III, the PIC can retrieve the power information of one meter locally within ($T_{Ethernet}+T_{Metering}$), which is less than 1 second. To finish the power information retrieval of four meters and report these data to the control center, it will take

$$T_{Retrieval\_PIC} = 4T_{Ethernet} + 4T_{Metering} + 0.5T_{3G} \quad (5)$$

Subtract (5) from (4), we get

$$\begin{aligned}
T_{save} &= T_{Retrieval} - T_{Retrieval\_PIC} \\
&= 4(T_{3G} + T_{Metering}) - (4T_{Ethernet} + 4T_{Metering} + 0.5T_{3G}) \\
&= 3.5T_{3G} - 4T_{Ethernet} \\
&\approx 3.5T_{3G}
\end{aligned} \quad (6)$$

Compared to the power information retrieval time of the original system, the control center could save $3.5T_{3G}$ in power information retrieval, which is 17.5 seconds in worst case. Therefore, the data pushing system is much faster than the original system.

### B. Server Waiting Time for Duty Cycle Change

The improvement of the system performance - by adjusting the waiting time of the server, $T_{waiting}$, has been discussed in [3]. Time $T_{waiting}$ starts after the server receives the successful return of the change in pilot signal duty cycle and ends before sending the power information request. Time $T_{waiting}$ can be expressed in (7),

$$T_{Waiting} > T_{EV} - (T_{3gUplink} + T_{Cloud}) \quad (7)$$

where $T_{Ev}$ stands for the time between the EV receiving the command of charging the current and settling down the current change. Note that $T_{EV}$ depends on the initial current $I_{init}$ and final current $I_{final}$. From the experiment in section III, $T_{Cloud}$ can be neglected compare with $T_{3G}$. Let's assume that $T_{3gUplink}$ is approximately half of $T_{3G}$, which is round trip time between the cloud and the charging station; therefore, we can rewrite (7) into (8):

$$T_{Waiting} > T_{EV} - 0.5T_{3G} \quad (8)$$

The maximum value of the $T_{EV}$ in [3] is 6 seconds and $T_{3G}$ falls in the range of 5 seconds in most cases in the experiments in section III. Therefore, the maximum value of $T_{waiting}$ could be 3.5 seconds on server for 3G case to cover most of the cases. With the further information of $I_{init}$ and $I_{final}$, the $T_{waiting}$ can be set to lower value rather than a fixed value to speed up the system's performance.

### C. Implementation of Local Charging Algorithm

Because the PIC can retrieve and save the power information locally in the smart charging station, it can locally implement charging algorithms to control the charging station at the local level. With less 3G traffic, the local charging algorithm could be more efficient than the charging algorithm on the server level. Nevertheless, considering the calculation power of the microprocessor, only certain simple charging algorithms, such as Round-Robin and Schedule Time, can be implemented in the microprocessor. Thus, with local charging algorithms implemented on the charging station level, the server will only need to select the mode of the charging algorithms of each charging station. This will save significant server computing resources. However, because of computing power, more complex charging and scheduling algorithms may still need to be implemented on the server. With local charging algorithms implemented, the control center could handle a larger smart charging system due to the reduction of traffic between the control center and the smart charging stations.

## VI. CONCLUSION

The smart EV charging system not only enhances the stability and reliability of the local power system but also provides an energy efficient, economical, and user friendly technology for charging EVs. In this paper, the Power Information Collector (PIC) is proposed, designed, and implemented in the smart EV charging station to reduce the response time to one quarter of original time. This will significantly decrease the response of the system. Besides the implementation of PIC, the distributions of the round trip, power information retrieval times are measured, presented, and discussed. The waiting time require for the server to change the duty cycle of the smart charging station can be further reduced by converting the system to a data pushing model. Furthermore, the implementation of local charging algorithms becomes possible because the local controller now has the power information and the relay status of the meters. Since the charging algorithm is embedded in the charging station, the traffic between the charging station and the control center can be further reduced; therefore, a faster response of the system can be achieved. With the smart charging algorithms implemented on both the control center and the local charging stations, our infrastructure would serve as one of the key components in the nation wide smart grid system to provide more economical, energy efficient scheme.

## VII. ACKNOWLEDGEMENT

This work has been sponsored in part by a grant from the LADWP/DOE fund 20699 & 20686, (Smart Grid Regional Demonstration Project). This material is based upon work supported by the United States Department of Energy under Award Number DE-OE000012 and the Los Angeles Department of Water and Power. Neither the United States Government nor any agency thereof, the Los Angeles Department of Water and Power, nor any of their employees make any warranty, express or implied, or assumes any legal liability or responsibility for the accuracy, completeness, or usefulness of any information, apparatus, product, or process disclosed, or represents that its use would not infringe privately owned rights.

Reference herein to any specific commercial product, process, or service by trade name, trademark, manufacturer, or otherwise does not necessarily constitute or imply its endorsement, recommendation, or favoring by the United States Government or any agency thereof. The views and opinions of authors expressed herein do not necessarily state or reflect those of the United States government or any agency thereof.

## BIOGRAPHIES

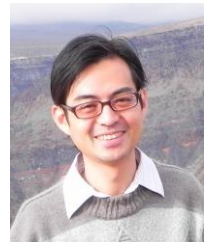

**Ching-Yen Chung** received his BS and MS degrees in Mechanical Engineering from National Taiwan University in Taiwan in 2002 and 2004. He had worked for the LITE-ON IT Corporation, an optical disc drive manufacturer, for five years. He received the university fellowship from UCLA graduate division in 2009 and 2012. He received the best paper bronze award in IEEE RFID TA conference in 2013. He is currently a PhD candidate in the Department of Mechanical Engineering, UCLA, developing the architecture for smart grid in WINMEC (Wireless Internet for Mobile Enterprise Consortium) Lab conducted by Professor Rajit Gadh.

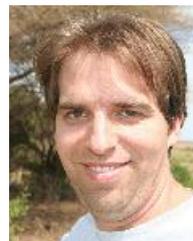

**Joshua Chynoweth** received his M.S. in Mechanical Engineering at UCLA in 2010, and his B.S. degree in Supply Chain Management from Arizona State University in 2001. He has lead multiple projects in three continents including being the founder and chief engineer for a film production company, constructing and running a microbiology lab for research in Tanzania, building a portable incubator for water quality assessments in Papua New Guinea. He is currently a PhD candidate in the UCLA Smart Gird Energy Research Center, developing an efficient feedback model for a smart micro-grid control system.

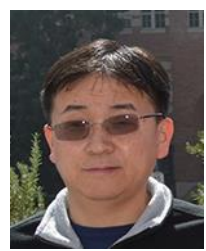

**Charlie Qiu** is presently a RESEARCHER at the Henry Samueli School of Engineering and Applied Science at University of California at Los Angeles. He has over 10 years of experience in RESEARCH AND DEVELOPMENT of software architectures, frameworks and solutions and has delivered multiple project solutions and software packages to the industry globally.

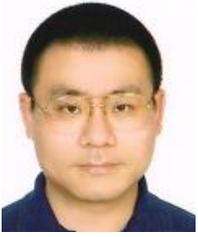 **Chi-Cheng Chu** received his PhD from University of Wisconsin – Madison on 2001 and BS form National Taiwan University on 1990. Dr. Chu is currently a project lead at Henry Samueli School of Engineering and Applied Science at University of California at Los Angeles. He is also the forum convener at UCLA WINMEC consortium. He has 2 patents attributed under his name and published more than 40 papers in professional engineering and scientific journals, books, and conference proceedings. He received one Best Paper Award in Excellence for Applied Research at 2004 Wireless Telecommunications Symposium.

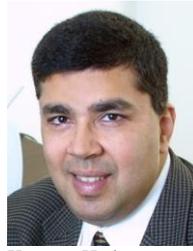 **Rajit Gadh** is a Professor at the Henry Samueli School of Engineering and Applied Science at UCLA, and the Founding Director of the UCLA Smart Grid Energy Research Center (SMERC) and the Wireless Internet for Mobile Enterprise Consortium (WINMEC). He has over 150 papers in journals, conferences and technical magazines. He has a Doctorate degree from Carnegie Mellon University (CMU), a Masters from Cornell University and a Bachelor's degree from IIT Kanpur. He has won several awards from NSF, SAE, IEEE, ASME, AT&T, Engineering Education Foundation, William Wong Fellowship award from University of Hong-Kong, ALCOA Science Support Scholar.